\documentstyle[prl,aps,epsf]{revtex}         
\tighten
\begin{document}
%
\twocolumn[\hsize\textwidth\columnwidth\hsize\csname @twocolumnfalse\endcsname 

\title{Observation of metastable states in spinor
Bose-Einstein condensates}
\author{H.-J. Miesner, D.M. Stamper-Kurn, J. Stenger, S. Inouye, A.P. 
Chikkatur, and W. Ketterle}
\address{Department of Physics and Research Laboratory of
Electronics, \\
Massachusetts Institute of Technology, Cambridge, MA 02139}
\date{November, 1998} \maketitle

\begin{abstract}
Bose-Einstein condensates have been prepared 
in long-lived metastable excited states.
Two complementary types of metastable states were observed.
The first is due to the 
immiscibility of multiple components in the condensate, and the 
second to 
local suppression of spin-relaxation collisions.
Relaxation via re-condensation of 
non-condensed atoms, spin 
relaxation, and quantum tunneling was observed.
These experiments were done with $F=1$ spinor Bose-Einstein 
condensates of sodium confined in an optical dipole trap.
\end{abstract}
\vskip1pc
]

Metastable states of matter, excited states which relax only slowly
to the ground state, are commonly encountered.
This slow relaxation often arises from the presence of free-energy barriers 
that prevent a system from directly evolving toward its ground state; 
if the thermal energy to overcome this barrier is not available, the 
metastable state may be long-lived.

Many properties of Bose-Einstein 
condensates in dilute atomic 
gases~\cite{ande:95,davi:95,brad:97,frie:98}
arise from metastability; 
indeed, such condensates are themselves 
metastable, since the true equilibrium state is a solid at 
these low temperatures.
Bose-Einstein condensates in gases with attractive interactions
(scattering length $a < 0$)~\cite{brad:97} are 
metastable against 
collapse due to a kinetic energy barrier~\cite{kaga:96}.
The persistence of rotations in condensates with 
repulsive interactions ($a>0$) hinges on whether vortices are 
metastable in singly-~\cite{rokh:97} or 
multiply-connected~\cite{rokh:97b,muel:98} geometries.
Similarly, dark solitons in restricted geometries are predicted to be 
long-lived~\cite{solitons} akin to the recently observed metastable 
states in superfluid $^{3}$He-B~\cite{back:98}.
Finally, Pu and Bigelow discussed spatial distributions 
of two-species condensates which are metastable due to mean-field 
repulsion between the two species~\cite{pu:98b}.

In this Letter, we report on the observation of two complementary types of
metastability in
$F = 1$ spinor Bose-Einstein condensates of sodium.
In one, a two-component condensate in the $|F=1, m_{F}=1, 0\rangle$ hyperfine 
states was stable in spin composition, but spontaneously formed a 
metastable spatial arrangement of spin domains.
In the other, a single component
$|m_{F}=0\rangle$ condensate
was metastable in spin composition with respect to
the development of $|m_{F}= \pm 1\rangle$ 
ground-state spin domains.
In both cases, the energy barriers which caused the metastability
(as low as 0.1 nK) were much smaller than the temperature 
of the gas (about 100 nK) which would suggest a rapid thermal 
relaxation.
However, this thermal relaxation was slowed considerably due to the high 
condensate fraction and the extreme diluteness of the non-condensed 
cloud.

Spinor Bose-Einstein condensates were created as in 
previous work~\cite{sten:98a}. 
Condensates in the $|F = 1, m_{F} = -1\rangle$ state were created in 
a magnetic trap~\cite{mewe:96a}
and then transfered to an optical dipole trap formed by a single 
infrared laser beam~\cite{stam:98a}.
The beam was weakly focused, producing cigar-shaped
traps with depths of 1 -- 2 $\mu$K, radial trap frequencies of about 
500 Hz,
and aspect ratios of about 70.
After transfer, Landau-Zener rf-sweeps placed
the optically trapped atoms into 
the desired initial spin states.
Condensates were then held for as long as 20 s
during which time the 
condensate numbers decayed from about $10^{6}$ to
$10^{5}$.
Condensate densities ranged from about 5 to $1 \times 10^{14} \, 
\text{cm}^{-3}$.

The spin-state distribution 
along the long axis of the optical trap was probed by a Stern-Gerlach 
spin separation combined with time-of-flight 
imaging~\cite{sten:98a}.
After a sudden switch-off of the optical trap, the condensate 
expanded primarily radially.
After 5 ms, a magnetic field gradient of
several G/cm separated the different spin-state populations
without distorting them.
Axial striations observed in time-of-flight images therefore reflected
the spatial structure of the trapped condensate.
After a total time-of-flight of 15 -- 30 ms, the atoms were 
optically pumped to the $|F = 2, m_{F} = 2\rangle$ hyperfine state,
and imaged with resonant light.

For our first set of experiments, a condensate in a
superposition of the $|F = 1, m_{F} = 1, 0\rangle$ 
hyperfine 
states was prepared by first transferring all the atoms to the 
$|m_{F} = 1\rangle$ state
with an adiabatic rf-sweep at low 
field (1 G), and then transferring half of the population to the
$|m_{F} = 0\rangle$ state with a 
non-adiabatic sweep at high field (15 G).
The sample then evolved freely in a 15 G 
bias field which prevented the formation of $|m_{F} = -1\rangle$ spin 
domains due to the large quadratic Zeeman energy barrier~\cite{sten:98a}.

Following their preparation, the two spin components
began to separate radially (Fig. 1, sideways in images). 
Within 50 ms, axial density striations were observed, indicating 
the formation of $|m_{F} = 1\rangle$ and $|m_{F} = 0 \rangle$ spin domains.
The striations were initially angled due to radial excitations
which tilted the domain walls at various angles to the long 
axis of the cloud.
After about 100 ms, the striations became horizontal, 
indicating the dissipation of radial motion and the orientation of 
the domain walls perpendicular to the axis of the cloud.
The spin domains reached typical sizes of
$40 \pm 15 \, \mu$m,
slightly larger than their size after 50 ms. 
Thereafter, the clouds were essentially unchanged -- only the total 
number of atoms diminished due to three-body losses.
Matching the images of the $|m_{F} = 1\rangle$ and $|m_{F} = 0 \rangle$
spin distributions showed that these long-lived states consist of 
persistent, alternating $|m_{F} = 1\rangle$ and $|m_{F} = 0 \rangle$ 
spin domains.

Phase-separation in two-component condensate mixtures is predicted 
by mean-field
theory~\cite{cols:78,ho:96,
esry:97,chui:98,gold:97}.
The mean-field interaction energy of such condensates is given 
by $2 \pi 
\hbar^{2}/ m \times  (n_{a}^{2} a_{a} + n_{b}^{2} a_{b} + 2 n_{a} n_{b}
a_{ab})$ where $m$ is the common atomic mass, $n_{a}$ and $n_{b}$ are the 
densities of each of the components, $a_{a}$ and $a_{b}$ are the 
same-species scattering lengths, and $a_{ab}$ is the scattering length 
for interspecies collisions.
The mean-field energy is minimized by phase-separation when the 
scattering lengths obey the relation $a_{ab} > \sqrt{a_{a} 
a_{b}}$.
At low magnetic fields, the scattering lengths in the $F = 1$ spinor 
system are determined by $a_{\normalsize F_{tot}=0}$ and
$a_{\normalsize F_{tot}=2}$ which describe collision between atoms 
with total angular momentum $F_{tot} = 0, 2$~\cite{ho:98}.
Defining $\bar{a} = (2 a_{\normalsize F_{tot}=2} + a_{\normalsize 
F_{tot}=0}) / 3$ and $\Delta a = (a_{\normalsize F_{tot}=2} - a_{\normalsize 
F_{tot}=0}) / 3$, the scattering lengths for the
$|m_{F} = 1 , 0\rangle$ two-component system are given by
$a_{0} = \bar{a}$, and $a_{1} = a_{01} = \bar{a} + \Delta a$.
Since $\Delta a$ was measured to be $3.5 \pm 1.5 \, 
\text{bohr} > 0$~\cite{sten:98a}, 
$a_{01} > \sqrt{a_{0}a_{1}}$ and the components phase-separate.
Interestingly, this phase-separation should not occur in the 
non-condensed cloud because the same-species mean-field interaction 
energies are doubled due to exchange terms.

The tendency of a 
two-component mixture to phase-separate implies the presence of 
imaginary frequencies for out-of-phase collective excitations of the overlapping
components (spin waves)~\cite{gold:97,grah:98}.
This means that the condensates are unstable, and slight 
perturbations grow exponentially.
In our experiment, density
and magnetic field inhomogeneities may yield slight perturbations in the
initial state, corresponding to an initial distribution of spin wave 
excitations.
The spontaneous formation of spin domains thus 
constitute an observation of spin waves with imaginary frequencies 
at wavelengths of 10 -- 50 $\mu$m, corresponding 
to the size of the spin domains which are intially formed.

The many-domain spin distribution is a 
macroscopically occupied excited state;
the ground state of the two-component system would contain only one 
domain each of $|m_{F} = 0\rangle$ and $|m_{F} = 1\rangle$ atoms, 
minimizing the surface energy of the domain walls~\cite{ao:98}.
For the many-domain state to decay directly to the ground state, the 
two condensate components would have to either overlap, or else
pass by each other without overlapping.
Such motion is energetically forbidden: the former
due to the mean-field energy (typically 50 Hz or 2.5 nK~\cite{sten:98a}),
and the latter due to the large kinetic energy required to 
vary the condensate wavefunction radially 
(the condensate is about 5 $\mu$m wide, yielding a 
kinetic energy barrier of 50 -- 100 Hz). 
For this reason, the excited states we observe are metastable.
In comparison, experiments on two-component mixtures of $^{87}$Rb 
showed no metastability, due in part to the nearly spherical 
traps (aspect ratio of $1 / \sqrt{8}$) which were used~\cite{hall:98a}.

The typical size of the metastable spin domains can be understood by 
considering the width of the domain walls 
($\simeq$ 3 $\mu$m~\cite{minfootnote}).
Spin domains smaller than this width
do not prevent the tunneling of unlike-spin atoms 
through the domain, and thus small spin domains are free to migrate 
through the cloud and coalesce with other like-spin domains.
Spin domains which are several times larger than the boundary width
prevent the tunneling of atoms across the domain, 
and are thus stable.
Metastable domains shorter than $\approx$ 25 
$\mu$m were not observed,
although the above description suggests that domains as short 
as $\approx$ 10 $\mu$m could be stable.
This may be due to either an axial smearing of small domains during 
their free expansion prior to imaging, or perhaps the destabilization 
of small domains by the initial radial excitations in the trapped 
cloud.

The spontaneously formed metastable states were extremely long lived. 
In the absence of an externally applied magnetic field gradient, the 
states persisted for at least 20 seconds, by which time the 
condensate number had dropped so low that we were unable to properly 
assess the time-of-flight images.
The lifetime of the metastable state was shortened by applying a 
magnetic field gradient along the axis of the trap.
At field gradients of 0.1 G/cm or greater, the 
metastable states decayed to the ground state within experimentally 
accessible times of 10 seconds or less.

To study this decay with a well-defined initial state, we first 
created a metastable two-domain system.
A strong magnetic field gradient was applied, causing the 
many spin domains to collapse into the ground state of two domains 
with the $|m_{F} = 1\rangle$ domain at the high-field end of the cloud.
A weaker field gradient (0.1 G/cm) was then applied in the 
opposite direction, which energetically favored the rearragement of 
the spin domains at opposite ends of the cloud.
In spite of this gradient, the two-domain system was metastable (Fig.\ 2).
It decayed slowly to equilibrium via two processes: a slow 
process which dominated for the first 12 s,
and a fast process (between 12 s and 13 s) which suddenly depleted
the metastable domains.

The slow process caused a gradual increase in the number of atoms
in the two ground state spin domains, which saturated at
about $5 \times 10^{4}$ atoms in each.
The rate of accumulation did not depend strongly on 
condensate density,
which was varied by decreasing the total number of atoms 
before creating the metastable state.
Thus, the slow process is probably due to re-condensation of the 
dilute thermal cloud into ground state spin domains, driven by a 
difference $\Delta \mu$ between the chemical potential of the thermal 
cloud and the energy of the ground state.

Studies of the formation of one-component Bose condensates in harmonic 
traps~\cite{mies:98,formation}
considered the case $T \gg \mu$ where $T$ is the 
temperature of the cloud, and $\mu$ is the chemical potential of the 
condensate ($\mu > 0$ due to interactions).
For optically trapped Bose condensates, the temperature is 
about 1/10 of the optical trap depth due to evaporation, i.e.\ $T 
\approx 100$ nK~\cite{stam:98a}.
For typical densities of $3 \times 10^{14} \, \text{cm}^{-3}$, $\mu = 
220 \, \text{nK} > T$.
Thus, previous studies of condensate formation may not apply to these 
metastable states.

The sudden decay to the ground state after the slow relaxation 
indicated the onset of a faster relaxation mechanism, which 
occurred once the decaying condensate had reached a critical density.
The dependence of this critical density on the applied field gradient 
suggests the onset of macroscopic quantum tunneling across the 
metastable spin domains.
Such behaviour will be reported in greater detail elsewhere.

The metastable states discussed so far involved multi-component 
mixtures stable in composition, but metastable in their 
spatial distribution.
Conversely, we identified another form of metastability in which a 
single component condensate was stable in its spatial distribution, 
but metastable in its hyperfine composition.
This was observed in
spinor condensates with 
an overall spin projection 
$\langle F_{z} \rangle = 0$ along the magnetic field axis~\cite{sten:98a}.
Condensates were prepared using Landau-Zener rf-sweeps at high 
fields (30 G) in one of two ways: either the entire 
trapped condensate was placed in the $|m_{F} = 0\rangle$ state by a single 
rf-sweep, or the condensate was placed in a 50-50 mixture of the 
$|m_{F}= +1\rangle$ and $|m_{F} = -1\rangle$ states using two rf-sweeps.
The condensates were then allowed to reach the ground state at a  
magnetic bias field $B_{0}$ and axial field gradient $B'$.

The evolution to equilibrium from the two starting conditions was 
quite different (Fig.\ 3).
When starting from the $|m_{F} = 0\rangle$
state, the condensate remained in that state for about
three seconds before 
evolving over the next few seconds to the ground state.
On the other hand, starting from an equal mixture of $|m_{F} = +1\rangle$ 
and $|m_{F} = -1\rangle$ atoms, the fraction of atoms in the $|m_{F} = 
0\rangle$ state grew without delay, arriving at equilibrium within 
just 200 ms.

This difference can be understood by considering 
a spin-relaxation collision, in which two $|m_{F} = 0\rangle$ atoms collide to 
produce a $|m_{F} = 1 \rangle$ and a $|m_{F} = -1 \rangle$ atom.
In the presence of a magnetic field $B_{0}$, quadratic Zeeman shifts 
cause the energy of the two $|m_{F} = 0\rangle$ atoms to be lower 
than the $|m_{F} = 1 \rangle$ and $|m_{F} = -1 \rangle$ atoms 
by $2 \times 390 \, \text{Hz} 
\times (B_{0} / \text{G})^{2}$.
Due to this activation energy, condensate atoms in the $|m_{F} = 
0\rangle$ state cannot undergo spin-relaxation collisions.
Thus, even though the creation of $|m_{F} = 1\rangle$ and $|m_{F} = -1 \rangle$
spin domains at the ends of the condensate
is energetically favored {\it globally} in the presence of a magnetic 
field gradient, the $|m_{F} = 0 \rangle$ condensate cannot overcome the 
{\it local} energy barrier for spin-relaxation.
On the other hand, 
condensate atoms in the $|m_{F} = 1\rangle$ and $|m_{F} = -1\rangle$ states 
can directly lower their energy through such collisions, and 
equilibrate quickly.

The decay of the metastable $|m_{F} = 0 \rangle$ state
was studied by varying the 
bias field $B_{0}$, the field gradient $B'$, and the trap depth $U$.
Evolution curves under various conditions (Fig.\ 4) were characterized 
by a delay time $T_{d}$ before the fraction of condensate atoms in the 
$|m_{F}=0\rangle$ state decayed exponentially to equilibrium with a 
relaxation time constant $\tau$.
Increasing the bias field $B_{0}$ did not affect 
$\tau$ significantly, but increased the
delay time from $T_{d} \simeq 1.9$ s at small bias fields 
(55 mG) to  $T_{d} \simeq 3.6$ s for larger fields (250 mG).
Lowering the trap depth from 2 $\mu$K to 1.2 $\mu$K increased both 
$T_{d}$ from about 0.6 to 4 s,
and $\tau$ from about 3.8 to 6.6 s. 
Increasing the magnetic field gradient only induced a weak lowering 
of $\tau$, and $T_{d}$ was not affected significantly.

The delay time $T_{d}$ should reflect the time-scale
for the gradual 
accumulation of $|m_{F} = \pm 1\rangle$ atoms in the thermal cloud 
before they reach a critical density and condense into the ground-state spin domains.
Increasing the bias field increases the 
quadratic Zeeman energy barrier for spin-relaxation, lengthening 
$T_{d}$.
However, the clear variation of $T_{d}$ as the energy barrier 
is changed from 0.1 nK at 55 mG to 2.5 nK at 250 mG, energies much 
smaller than the temperature of $T \approx 100$ nK, cannot be 
explained on the basis of thermal spin-relaxation collisions.
Further study of these relaxation timescales is warranted.
	
In conclusion, we have identified
a novel form of metastability in Bose-Einstein condensates which 
occurs at temperatures much higher than the energy barrier for 
evolution to the ground state.
This metastability arises due to the high condensate fraction which 
dramatically slows down thermal relaxation.
This offers the opportunity to study the kinetics of condensation and 
of the dissolution of metastable condensates in 
``slow motion,'' on times scales much longer than collision times and 
of periods of collective excitations.
Furthermore, we have observed spontaneous phase-separation in a 
two-component mixture in accordance with the predictions of mean-field 
theory.

This work was supported by the Office of Naval Research, NSF, Joint Services
Electronics Program (ARO), NASA, and the David and Lucile Packard Foundation. 
A.~P.~C.  acknowledges additional support from the NSF, D.~M.~S.~-K. from JSEP,
and J.~S. from the Alexander von Humboldt-Foundation.

\bibliographystyle{prsty}

\begin{thebibliography}{99}
\bibitem{minfootnote}
Approximating the two condensate wavefunctions in the boundary 
region ($0 \leq z \leq L$) to be
$\psi_{1} = \sqrt{n_{1}} \cos \pi z / 2 L$,
$\psi_{0} = \sqrt{n_{0}} \sin \pi z / 2 L$,
and assuming equal pressure ($n_{1}^{2} a_{1} = n_{0}^{2} a_{0}$) 
across the boundary of length $L$, the kinetic and mean-field (see 
text) energies are minimized at
$L^{2} = \pi \xi_{1}^{2} \frac{a_{1} + \sqrt{a_{1} a_{0}}}{a_{0 1} - 
\sqrt{a_{1} a_{0}}}$.
Here $\xi_{1} = (8 \pi n_{1} a_{1})^{-1/2}$.
For $n_{1} \simeq 3 \times 10^{14}$ cm$^{-3}$, and $a_{1} = 52$ bohr,
$L \simeq 3 \, \mu$m.
\end{thebibliography}

\begin{figure*}
\begin{caption}
{Spontaneous formation of metastable states in a 15 G bias field.
Condensates were probed at various times after preparing an overlapping 
$|m_{F} = 1, 0\rangle$ mixed condensate.
The $|m_{F} = 1\rangle$ (top) and $|m_{F} = 0\rangle$ (bottom) states 
were separated during their free expansion.
Striations reflect the presence of alternating $\approx 40 \, 
\mu$m spin domains which form 50 ms after state preparation.
High resolution ($\approx 10 \, \mu$m) was achieved by only optically 
pumping a 200 $\mu$m slice of the clouds.  The height of each image is 
1.3 mm.}
\end{caption}
\label{fig1}
\end{figure*}
\begin{figure}
\epsfxsize = 80mm
\centerline{\epsfbox[0 0 424 140]{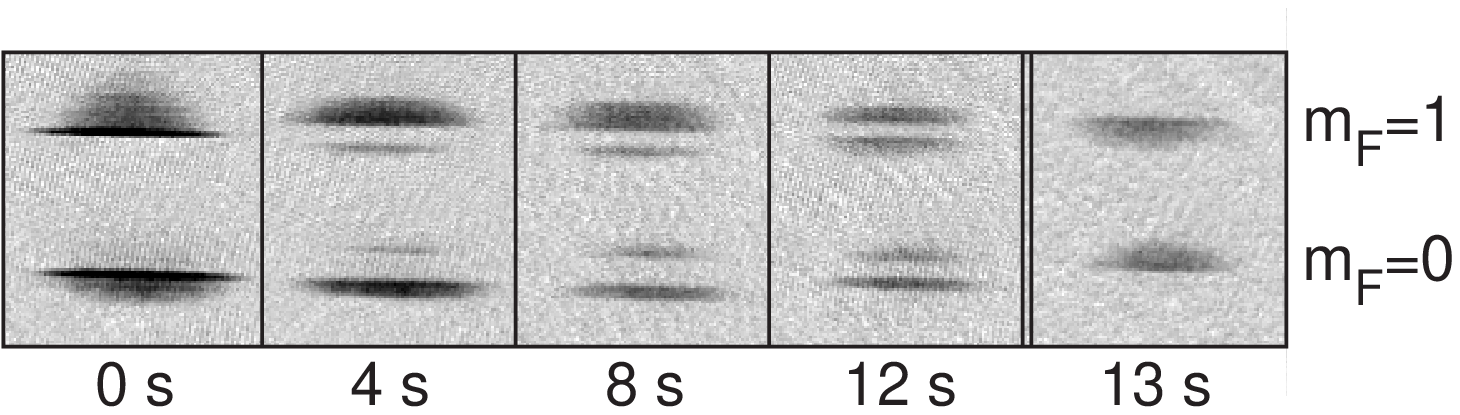}}
\begin{caption}
{Relaxation of metastable states in 
a 15 G bias field and a 0.1 G/cm gradient.
Images show metastable (outer) and 
ground state (inner) spin domains of $|m_{F} = 1\rangle$ and $|m_{F} = 
0\rangle$ atoms, probed at various times after state preparation.
The metastable domains decayed slowly for 12 s 
before tunneling quickly to the ground state.
The height of each image is 1.3 mm.
}
\end{caption}
\label{fig2}
\end{figure}

\begin{figure}
\epsfxsize = 70mm
\centerline{\epsfbox[0 0 373 201]{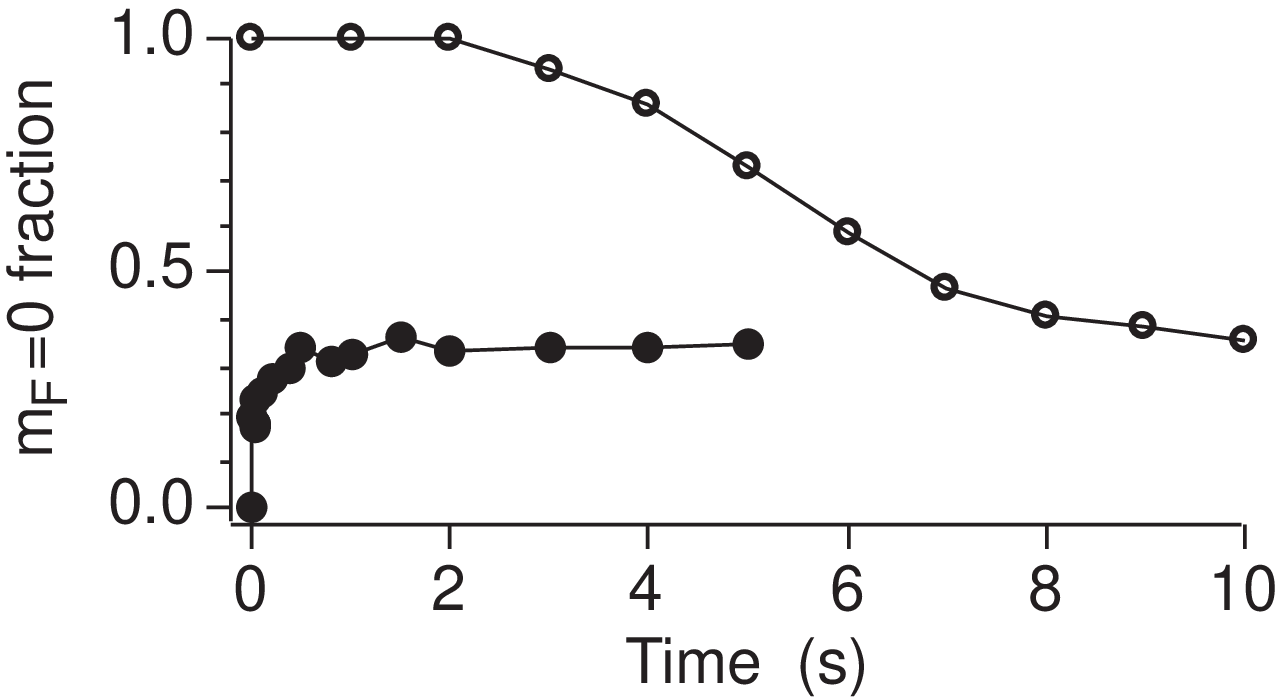}}
\label{fig3}
\caption{Metastability of the pure $|m_{F} = 0\rangle$ state in the 
presence of a magnetic bias field (250 mG), and gradient (44 mG/cm).
The evolution toward equilibrium of a pure $|m_{F} = 0 \rangle$ 
condensate (open symbols),
and a mixture of $|m_{F} = 1\rangle$ and $|m_{F} = -1 \rangle$ (closed 
symbols) is 
shown by plotting the observed fraction of atoms in the $|m_{F} = 0 
\rangle$ state vs.\ dwell time in the optical trap.
}
\end{figure}

\begin{figure}
\epsfxsize = 60mm
\centerline{\epsfbox[0 0 354 309]{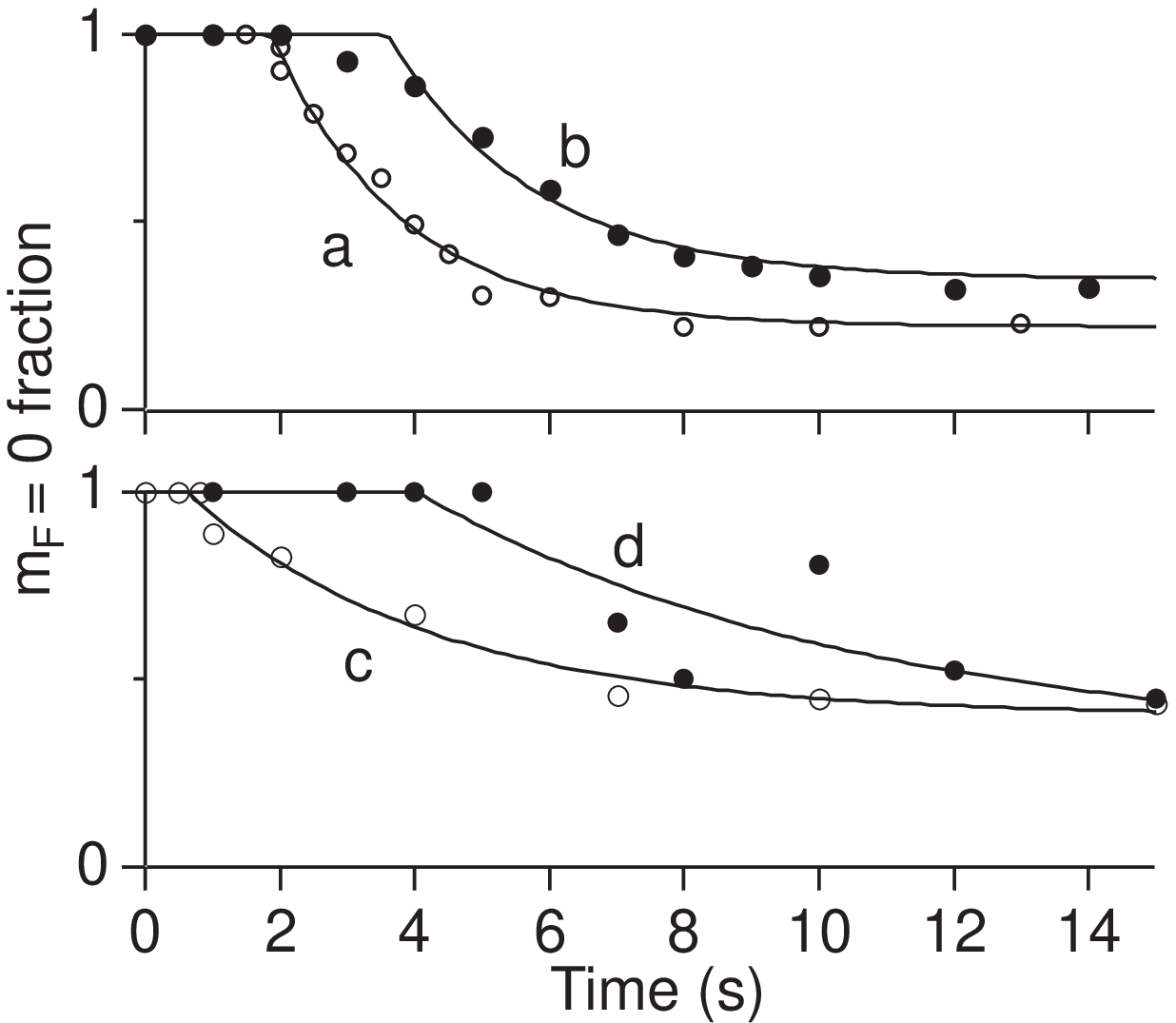}}
\label{fig4}
\caption{Decay of $|m_{F} = 0\rangle$ metastable states.
The lines are exponential decay fits to the $|m_{F} = 0\rangle$ 
fraction after a delay time $T_{d}$.
Increasing the bias field (top) increased $T_{d}$.  Lowering the 
temperature of the gas (bottom) increased both $T_{d}$ and $\tau$.
The magnetic bias field $B_{0}$, the gradient $B'$, and the trap depth 
$U$ for each curve are: a) 55 mG, 44 mG/cm, 2 $\mu\text{K}$;
b) 250 mG, 44 mG/cm, 2 $\mu\text{K}$; 
c) 20 mG, 11 mG/cm, 2 $\mu\text{K}$; 
and d) 20 mG, 11 mG/cm, 1.2 $\mu\text{K}$.
}
\end{figure}

\end{document}